\documentclass{aa}
\usepackage{graphicx,float}

\def \sax {BeppoSAX}
\def \src {XTE{\thinspace}J0421+560}
\def \star {CI{\thinspace}Cam}
\def \degmark{^\circ}
\def \nh {N${\rm _H}$}

\def \hcm {\hbox {\ifmmode $ atom cm$^{-2}\else atom cm$^{-2}$\fi}}
\def \arcmin {\hbox{$^\prime$}}

\def\approxgt{\mathrel{\hbox{\rlap{\lower.55ex \hbox {$\sim$}}
        \kern-.3em \raise.4ex \hbox{$>$}}}}
\def\approxlt{\mathrel{\hbox{\rlap{\lower.55ex \hbox {$\sim$}}
        \kern-.3em \raise.4ex \hbox{$<$}}}}
\newcommand{\mc}{\multicolumn}

\begin{document}

\thesaurus{(08.02.05; 08.09.2; 08.14.2; 13.25.5)}  

\title{Unusual quiescent X-ray activity from \src\ (CI{\thinspace}Cam)}

\author{A.N. Parmar\inst{1}
         \and T. Belloni\inst{2}
         \and M. Orlandini\inst{3}
         \and D. Dal Fiume\inst{3}
         \and A. Orr\inst{1}
         \and N. Masetti\inst{3}
}
\offprints{A.N.Parmar (aparmar@astro.estec. esa.nl)}

\institute{
       Astrophysics Division, Space Science Department of ESA, ESTEC,
       Postbus 299, NL-2200 AG Noordwijk, The Netherlands
\and
       Osservatorio Astronomico di Brera, via Bianchi 46, 23807 Merate,
       Italy
\and
       Istituto Tecnologie e Studio Radiazioni Extraterrestri, CNR, 
       Via Gobetti 101, 40129 Bologna, Italy
}
\date{Received: 2000 May 22; Accepted: 2000 May 30}

\maketitle

\markboth{\src\ in quiescence}{\src\ in quiescence} 

\begin{abstract}
We report on BeppoSAX observations of the X-ray transient \src\
in quiescence 156, 541, and $\sim$690 days after the 
maximum of the 1998 April outburst.
In the first observation the source was soft with a power-law
photon index, $\alpha$, of $4.0 \pm ^{1.9} _{0.9}$
and absorption, ${\rm N_H}$, of 
$(1.1 \pm  ^{4.9} _{1.1}) \times 10^{21}$~atom~cm$^{-2}$.
In the second observation, the source brightened by a factor
$\sim$15 in the 1--10~keV energy range, became significantly
harder with $\alpha = 1.86 \pm ^{0.27} _{0.32}$ and was strongly absorbed with
${\rm N_H = (4.0 \pm 0.8) \times 10^{23}~atom~cm^{-2}}$. 
There is evidence for a narrow emission line in both spectra at $\sim$7~keV.
In the third
observation, the source had faded by a factor $\approxgt$8 from the
previous observation to below the BeppoSAX detection level.
It is possible that these variations result from orbital motion
of a compact object around the B[e] star companion with the
intense, absorbed, spectrum arising during passage through 
dense circumstellar material. If this is the case,
the system may be continuing to exhibit periodic activity.

\end{abstract}

\section{Introduction}
\label{sect:intro}

On 1998 March 31 a soft X-ray transient (\src) was detected by the
All-Sky Monitor onboard R-XTE (Smith et al. \cite{s:98}).
The source brightened rapidly, reaching $\sim$2~Crab after a few hours
and then quickly decayed with an initial $e$-folding time of
only 0.5~day. The outburst was observed by CGRO 
(Paciesas \& Fishman \cite{pf:98}), 
R-XTE (Belloni et al. \cite{b:99}), ASCA (Ueda et al. \cite{u:98}) 
and BeppoSAX (Frontera et al. \cite{f:98}; Orr et al. \cite{o:98}).
This was the fastest rise and decay of any outburst from a 
soft X-ray transient (see e.g., Chen et al. \cite{c:97}).
The outburst X-ray spectra are complex
and cannot be fit by any of the models usually applied to
soft X-ray transients. 
A two-temperature bremsstrahlung model was used to describe the 
spectra of the ASCA and two BeppoSAX outburst observations. 
Both BeppoSAX spectra included line
features identified with O, Ne/Fe-L, Si, S, Ca and Fe-K. During the
second observation the O and
Ne/Fe-L line energies decreased smoothly by $\sim$9\%, while the other line 
energies remained unchanged (Orr et al. \cite{o:98}). 
No bursts, pulsations, or quasi-periodic
pulsations have been detected, and so the origin and nature of the
X-ray emission remains uncertain. 

Radio and optical observations identified \src\ with
a B[e] star \star\ (MWC\thinspace84).
Radio observations of \src\ revealed a slow
($\sim$1000~m~s$^{-1}$), shell-like motion (Hjellming et al. \cite{h:98})
as well as the presence of SS\thinspace443-like jets 
(Hjellming \& Mioduszewski \cite{hm:98a}, \cite{hm:98b}) with a 
velocity of $0.15~c$.
Spectroscopic observations by 
Wagner \& Starrfield (\cite{w:98})
revealed a rich emission line spectrum with He~{\sc ii} in emission.
These features were absent in previous spectra (Downes \cite{d:84}).
The class of B[e] stars include
many objects of different types and evolutionary status
(e.g., Lamers et~al. \cite{l:98}). 
They differ from classical Be stars in showing
a pronounced IR excess due to emission from warm dust, rather than
the free-free and free-bound emission from the gaseous envelopes of
classical Be stars. Near IR spectroscopy of \star\ by Clark et al. 
(\cite{c:00})
reveals a complex circumstellar environment with a highly ionized
region (presumably near the compact object) responsible for the 
He~{\sc ii} emission embedded in a dense stellar wind. The
circumstellar envelope includes cold dense regions which may by
located in a disk, or the result of discrete mass ejections.

The distance to \star\ is quite uncertain. Hjellming (private
communication) infers a distance of $1.0 \pm 0.2$~kpc from
an observation of the 21~cm H~{\sc i} absorption profile. 
Based on the optical properties of the source Zorec (\cite{z:98}) 
and Clark et al. (\cite{c:00}) estimate distances 
of 1.75~kpc and $\sim$0.6--2.0~kpc, respectively.  
This uncertainty in distance
is important, since at 2~kpc the peak outburst luminosity
of several $10^{37}$~erg~s$^{-1}$ is high enough to exclude accretion
onto a white dwarf as the origin of the (assumed isotropic) X-ray emission. If the source 
is at 1~kpc, then it is not. 

In quiescence the X-ray luminosity of soft X-ray transients
appears to depend on the nature of the compact object. From a 
systematic study of 13 soft X-ray transients, Asai et al. (\cite{a:98})
find that the quiescent luminosity of systems containing neutron stars
is $\sim$$10^{32-33}$~erg~s$^{-1}$, whereas those of systems
containing blackholes are systematically lower with most of them
having upper limits in the range $\approxlt$$10^{32}$~erg~s$^{-1}$.
In order to investigate the nature of any quiescent emission, 
\src\ was observed by BeppoSAX in 1998 September,
156 days after the outburst maximum. A faint source at a position
consistent with \src\ was detected (Orlandini et al. \cite{o:00}). 
The spectrum can be fit with the same two temperature 
(kT$_1$, kT$_2$) model as during
the outburst with the same value of kT$_1$ as during the second
outburst observation and a lower value of kT$_2$.
We report here also on two further BeppoSAX observations of \src\
in quiescence, made 541 and $\sim$690 days after outburst maximum.
During the first observation, in 1999 September,
an absorbed, hard source was detected, while
during the second, in 2000 February, the source intensity was below
the detection threshold. 

\begin{table*}
\caption{BeppoSAX observing log of \src\ in quiescence. (There were also 2
BeppoSAX observations in 1998 April during the outburst.)
The third quiescent observation
consists of two pointings separated by 5 days which are analyzed together.
Uncertainties and upper limits are given at 3$\sigma$ confidence. The
flux estimates are discussed in Sect.~\ref{sect:spectrum}}
\begin{tabular}{llrrrccr}
\hline\noalign{\smallskip}
Obs & \mc{2}{c}{Observation} & \mc{2}{c}{Exposure}& \mc{2}{c}{Count rate}
& \mc{1}{c}{Flux}\\
       &  \mc{1}{c}{Start} & \mc{1}{c}{End} & \mc{1}{c}{LECS} & \mc{1}{c}{MECS}
       &\mc{1}{c}{LECS} & \mc{1}{c}{MECS} &\mc{1}{c}{(1--10~keV)} \\
       & (yr~mn~dy~hr:mn) & (mn~dy~hr:mn) & (ks) &(ks)
&(0.1--8 keV; s$^{-1}$)& (2--10~keV; s$^{-1}$) &(erg~cm$^{-2}$~s$^{-1}$) \\
\noalign {\smallskip}
\hline\noalign {\smallskip}
1 & 1998~Sep~03 14:19 & Sep~04 14:19 & 19.1 & 44.8 & $0.0033 \pm 0.0006$ &
$0.0015 \pm 0.0003$ & $4.5 \times 10^{-13}$ \\
2 & 1999~Sep~23 11:17 & Sep~25 15:07 & 30.8 & 60.7 & $0.0024 \pm 0.0010$  &
$0.0078 \pm 0.0004 $& $7.6 \times 10^{-12}$ \\
3 & 2000~Feb~20 10:05 & Feb~21 20:41 & 13.0 & 66.5 & \llap{$<$}0.0020   &
\llap{$<$}0.00092 & $<$$9 \times 10^{-13}$\\
 & 2000~Feb~25 11:03 & Feb~26 05:42 &      &      &                    &
                     \\
\noalign {\smallskip}
\hline
\end{tabular}
\label{tab:observing_log}
\end{table*}

\section{Observations and results}
\label{sect:obs}

Results from the imaging Low-Energy Concentrator Spectrometer (LECS;
0.1--10~keV; Parmar et al. \cite{p:97}) and Medium-Energy Concentrator
Spectrometer (MECS; 1.8--10~keV; Boella et al. \cite{b:97})
on-board BeppoSAX are presented. Due to the faintness of the source
\src\ was not detected by the non-imaging high-energy
instruments.
The region of sky containing \src\ was observed three times by BeppoSAX
following the end of the 1998 April outburst
(see Table~\ref{tab:observing_log}). The results of the first
of these observations are reported in Orlandini 
et al. (\cite{o:00}) and we include these data here for
completeness. As usual, good
data were selected from intervals when the elevation angle
above the Earth's limb was $>$$4^{\circ}$ and when the instrument
configurations were nominal, using the SAXDAS 2.0.0 data analysis package.
LECS and MECS data were extracted centered on the position of \src\ 
using radii of 4\arcmin\ and 2\arcmin, respectively. 

A source with a MECS 2--10~keV count rate of 0.0078~s$^{-1}$ was 
detected at a position consistent with \src\ during the 1999 
observation. This value can be compared to the MECS 2--10~keV 
count rate during the 1998 September observation of 0.0015~s$^{-1}$,
indicating that the source had brightened substantially.
There is no evidence for any variability during the 1999 observation
with a 3$\sigma$ limit to rms fractional variability of 7.1\%.
During the 2000 February
observation, the source was not detected with a 3$\sigma$ upper limit
to the 2--10~keV MECS count rate of 0.00092~s$^{-1}$. 
We note that a preliminary comparison of optical (4000--8000 \AA) spectra 
acquired simultaneously with the 1999 and 2000 BeppoSAX observations 
(Bartolini et al., in preparation), as well as with the
optical spectra reported by Orlandini et al. (\cite{o:00}), shows no 
obvious differences in the relative strengths
of the Balmer and He~{\sc i} emission lines.

The usual method of background subtraction for the LECS and the
MECS is to use deep exposures of ``standard'' high galactic latitude fields.
Since the X-ray sky background is spatially structured, especially at
low energies (see e.g., Snowden et al. \cite{s:95}), this may not 
be appropriate for \src\ (l, b = $149\degmark, +4.1\degmark$).
To investigate this effect,
LECS background spectra were produced using the annulus method 
described in Parmar et al. (\cite{p:99}), using the central 8\arcmin\ of 
the LECS and MECS fields of view during the 2000 observation (when
the source was not detected), and using standard files. The spectral fit 
results for the 1998 September and 1999 observations do not depend 
significantly on which method was used, and all quoted results used
the annulus method.

\begin{figure*}
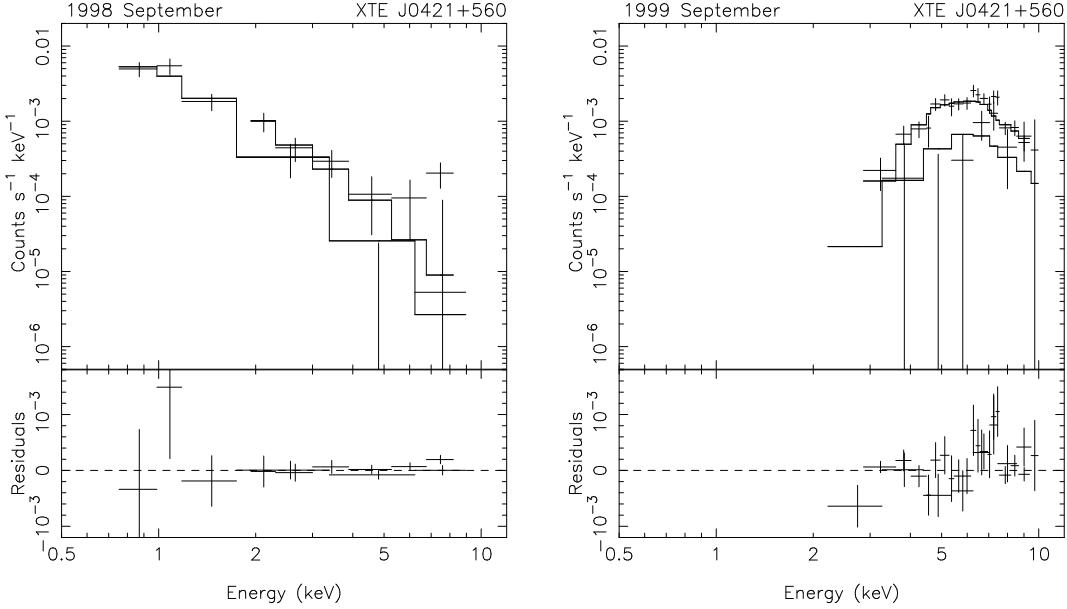

  \centerline{
  \hbox{\hspace{-1.0cm}
   \includegraphics[width=8.0cm,angle=-90]{ce231_fig1a.ps}
   \hspace{0.5cm}
   \includegraphics[width=8.0cm,angle=-90]{ce231_fig1b.ps}}}
  \caption[]{The observed LECS and MECS spectra of \src\ together with the
             best-fit power-law and Gaussian line models for the 1998 
             September (left panels) and 1999 (right panels) observations.
             In the lower panels the line normalization has been
             set to zero}
  \label{fig:spectra}
\end{figure*}

\section{Spectral fits}
\label{sect:spectrum}

In order to further investigate the nature of this brightening, 
LECS and MECS spectra for the 1998 September and 1999 observations
were extracted. They were rebinned to oversample the full
width half maximum of the energy resolution by
a factor 3 and to have additionally a minimum of 20 counts 
per bin to allow use of the $\chi^2$ statistic.
The photoelectric absorption
cross sections of Morisson \& McCammon (\cite{m:83}) were used throughout.
A factor, constrained to be within its usual 0.8--1.0 range,
was included in the spectral fit models to allow for the
normalization uncertainty between the LECS and MECS. 
The results of simple model fits,
including absorbed power-law,
thermal bremsstrahlung and collisionally ionized thermal equilibrium
plasma ({\sc mekal} in {\sc xspec}) models, are given in 
Table~\ref{tab:spectrum}. In the {\sc mekal} fits, the
abundance of the emitting material was fixed at the solar value.
Orlandini et al. (\cite{o:00}) report the detection of an emission
feature with an energy of $\sim$7~keV in the 1998 September spectrum.
We therefore included a power-law model together with a
narrow Gaussian emission feature in the fits.
The acceptable 
values of $\chi ^2$ indicate that more complex models are not
required.

Table~\ref{tab:spectrum} shows 
the very different spectral
shapes during the two observations. In 1998 September the source
could be modeled by a power-law with a photon index, $\alpha$,
of $4.0 \pm ^{1.9} _{0.9}$ and 
${\rm N_H}$ of $(1.1 \pm ^{4.9} _{1.1}) \times 10^{21}$~atom~cm$^{-2}$,
together with a narrow 
Gaussian line feature at $7.0 \pm ^{1.6} _{0.2}$~keV
for a $\chi^2$ of 3.3 for 6 degrees of freedom (dof). 
The same model also gives the best-fit to the 1999 spectrum, but now
$\alpha$ is $1.86 \pm ^{0.27} _{0.32}$ and ${\rm N_H}$ is
$(4.0 \pm 0.8)\times 10^{23}$~atom~cm$^{-2}$ for
a $\chi^2$ of 28.0 for 28 dof. The line energy is $7.3 \pm 0.2$~keV
and the equivalent width $620 \pm 350$~eV.
Fig.~\ref{fig:spectra}
shows the observed LECS and MECS spectra for the 2 observations
together with the best-fit (power-law and Gaussian line) model.
The contrast in spectral shape is striking. During the 1999
observation the source was not detected by the LECS below 2~keV, whereas
the majority of the emission was below this energy in the 1998 September
observation. The unabsorbed 1--10~keV fluxes given
in Table~\ref{tab:observing_log} are derived using the best-fit
parameters and by setting ${\rm N_H = 0}$. 
In the case of the 2000 observation, the upper
limit to the flux is derived assuming the absorption corrected 
spectral shape of the 1999 observation (when the source was
strongly absorbed). 
If the absorption corrected spectral shape
from the 1998 September observation is used, the 3$\sigma$ upper-limit
to any 1--10~keV flux is $2.8 \times 10^{-13}$~erg~cm$^{-2}$~s$^{-1}$.
This implies that the source had faded by a factor $\approxgt$8 in the 1--10~keV
energy range from the 1999 observation.

\begin{table}
\caption[]{\src\ fit results.
\nh\ is in units of $\rm {10^{22}}$ atom $\rm {cm^{-2}}$.
90\% confidence limits are given. For the {\sc mekal} fits
the abundance was fixed at the solar value. PL = Power-law}
\begin{flushleft}
\begin{tabular}{lcccr}
\hline\noalign{\smallskip}
Model & \hfil N$_{\rm {H}}$ \hfil & kT (keV) &$\alpha$ & $\chi^2$/dof \\
\noalign{\smallskip\hrule\smallskip}
Observation 1 \\
\quad Power-law &  $0.0 \pm ^{0.44} _{0.0}$ & \dots & $3.5 \pm ^{1.7} _{0.6}$  
& 10.3/8  \\
\quad Brems. & $0.0\pm ^{0.17} _{0.0}$ & $0.76 \pm ^{0.42} _{0.28}$ & 
\dots & 15.0/8\\
\quad {\sc mekal}    & $0.0 \pm ^{0.54} _{0.0}$ & $1.24 \pm ^{0.24} _{0.47} 
$ & \dots & 10.1/8\\
\quad PL + Line & $0.11 \pm ^{0.49} _{0.11}$ & \dots & $4.0 \pm ^{1.9}
_{0.9}$ & 3.3/6 \\
\noalign{\smallskip\hrule\smallskip}
Observation 2 \\
\quad Power-law & $41 \pm ^{15} _{4}$     & \dots & $1.72 \pm ^{0.80} _{0.38}$ 
& 37.3/30  \\
\quad Brems. & $41 \pm ^{13} _{11}$ & $>$6.4 & 
\dots & 37.0/30\\
\quad {\sc mekal}    & $35 \pm 6 $ & $>$14 & \dots & 34.2/30\\
\quad PL + Line & $40 \pm 8$ & \dots & $1.86 \pm ^{0.27} _{0.32}$ 
& 28.0/28 \\
\noalign{\smallskip\hrule}
\end{tabular}
\end{flushleft}
\label{tab:spectrum}
\end{table}

From an analysis of diffuse interstellar bands in the optical spectrum
of \star, Clark et al. (\cite{c:00}) estimate that the interstellar
E(B--V) is $0.65 \pm 0.20$. This implies ${\rm A_v = 2.0 \pm 0.6}$. 
Using the relation 
${\rm N_H [ cm^{-2}/A_v] = 1.79 \times 10^{21}}$ from
Predehl \& Schmitt (\cite{ps:95}) implies that observed X-ray absorption
should be not less than $(3.6 \pm 1.1)\times 10^{21}$~atom~cm$^{-2}$.
The best fit to the 1998 September spectrum gives an 
${\rm N_H}$ of 
$(1.1 \pm ^{4.9} _{1.1})\times 10^{21}$~atom~cm$^{-2}$, 
consistent with the predicted interstellar value. 
For a distance d in kpc,
the fluxes listed in Table~\ref{tab:observing_log} correspond
to 1--10~keV luminosities of $5.4 \times 10^{31}$, $9.1 \times 10^{32}$, 
and $<$$1 \times 10^{32}$~${\rm d^2 _{kpc}}$~erg~s$^{-1}$ for the
1998 September, 1999, and 2000 observations, respectively. We note
that the distance to \star\ is probably in the range 0.6 to 2.0~kpc
(see Sect.~\ref{sect:intro}). 

\section{Discussion}
\label{sect:discussion}

\src\ is a highly unusual X-ray transient due to (1) the high-mass nature
of its companion, (2) the extremely short outburst duration, (3) the
complex, time varying, line-rich outburst X-ray spectra to which must
now be added (4) the long duration, and highly variable quiescent emission.
In the 1998 September observation the source was soft with low
absorption ($1.1 \times 10^{21}$~atom~cm$^{-2}$) 
while during
the 1999 observation it had hardened and brightened by a 1--10~keV 
factor of $\sim$15 and
${\rm N_H}$ had increased to $4.0 \times 10^{23}$~atom~cm$^{-2}$.
In the 2000 observation \src\ was not detected indicating that
the source was a factor $\approxgt$8 times fainter than during the second
observation. Despite these variations in X-ray emission,
the main features of the optical spectrum of \star\ 
did not change appreciably.

The quiescent variability of soft X-ray transients has been poorly
studied. This is primarily because these objects are faint
in quiscence, with intensities
often close to the detection limit of current instruments, and therefore
difficult to observe (e.g., Verbunt et al. \cite{v:94}; 
Asai et al. \cite{a:98}; Campana \& Stella \cite{cs:00}).  
It is possible that the decreasing emission observed here represents
a long lived component of the original ($\sim$10~day duration) outburst.
In this case the brightening seen in 1999 may result from
some sort of re-flare. It is then unclear why the 1999 absorption
was so high, since near the peak of the 1998 April outburst the
absorption was only $\sim$$5 \times 10^{22}$~atom~cm$^{-2}$ which then 
decreased as the outburst decayed (Belloni et al. \cite{b:99}).
There is no evidence for similar long lived activity
from other soft X-ray transients (e.g., Chen et al. \cite{c:97}), 
although it is quite possible that similar low-level emission 
may have been missed due to a lack of suitable observations.

Even at the maximum likely distance of 2~kpc the upper-limit to the luminosity
during the 2000 observation of $<$$4 \times 10^{32}$~erg~s$^{-1}$
is more similar to those of quiescent soft X-ray transients containing
black holes than those containing neutron stars. However, the
comparison with other soft X-ray transients may be misleading
since it is unclear whether the accretion mechanism is the same in
all cases. The hard spectrum observed in 1999 means that at times
there may be sufficient hard X-ray photons to trigger the mass
loss instability of Hameury et al. (\cite{h:86})
discussed in the context of the 1998 September 
observation by Orlandini et al. (\cite{o:00}).

It is possible that the observed quiescent X-ray properties 
result from orbital motion. As the compact object passes through the
material located around the B[e] star the spectrum hardens,
becomes more intense, and the absorption increases. At some distance
from the B[e] star 
both the absorption and accretion rate decrease resulting in a 
spectrum similar to that observed in 1998 September. 
The non-detection in the 2000 observation may then result from an 
extremely low accretion rate far from the companion star.  
If this is the case, then the system may continue to be active
for some time and continued multi-waveband monitoring could be
important in helping elucidate the nature of the system.

\begin{acknowledgements}
The \sax\ satellite is a joint Italian-Dutch programme. 
\end{acknowledgements}

\end{document}